\documentclass[useAMS,usenatbib]{mn2e}
\usepackage{graphicx}
\usepackage{graphics}
\usepackage{url}
\usepackage{epstopdf}
\usepackage{color}
\usepackage{bm}
\usepackage{times}

\newcommand{\icarus}{Icarus}

\def\lsim{\mathrel{\hbox{\rlap{\hbox{\lower4pt\hbox{$\sim$}}}\hbox{$<$}}}}
\def\gsim{\mathrel{\hbox{\rlap{\hbox{\lower4pt\hbox{$\sim$}}}\hbox{$>$}}}}

\def\ft2pi{\frac{1}{\left(\sqrt{2\pi}\right)^{3}}}

\def\d3{\mbox{d}}

\begin{document}
%
%
%
%
%
%
%
%

\title[Polar ice on Mercury]{How thick are Mercury's polar water ice deposits?}

\author[V. R. Eke et al.]{
\parbox[t]\textwidth{
Vincent R. Eke$^1$,
David J. Lawrence$^2$,
Lu\'{\a i}s F. A. Teodoro$^3$}\\
\parbox[t]\textwidth{
$^1$Institute for Computational Cosmology, Department of Physics, Durham University, Science Laboratories, South Road, Durham DH1 3LE, U.K.\\
$^2$Johns Hopkins University Applied Physics Laboratory, Laurel, MD 20723, U.S.A.\\
$^3$BAER, Planetary Systems Branch, Space Science and Astrobiology Division, MS: 245-3, NASA Ames Research Center, Moffett Field, CA 94035-1000, U.S.A.}}

\maketitle

\begin{abstract}

An estimate is made of the thickness of the radar-bright deposits in
craters near to Mercury's north pole. To construct an objective set of
craters for this measurement, an automated crater finding algorithm is developed and
applied to a digital elevation model based on data
from the Mercury Laser Altimeter onboard the MESSENGER spacecraft.
This produces a catalogue of 663 craters with diameters exceeding
4\,km, northwards of latitude $+55^\circ$. A subset of 12 larger, well-sampled and
fresh polar craters are selected to search for correlations between
topography and radar same-sense backscatter cross-section. It is found
that the typical excess height associated with the radar-bright
regions within these fresh polar craters is ($50\pm35$)\,m. This puts
an approximate upper limit on the total polar water ice deposits on
Mercury of $\sim 3\times 10^{15}$\,kg.

\end{abstract}

\section{Introduction}

The presence of water in the inner solar system is vital for the
development of life. Violent collisions between planetary embryos are
thought to have built the rocky planets \citep{morbi12}, and the associated high
energies and temperatures would not have been conducive for volatile
molecules such as water. The source of Earth's water is therefore of
considerable interest. Measurements of deuterium/hydrogen ratios of solar system
bodies \citep{hartogh11,alexander12} combined with models of accretion onto the Earth
\citep{morbi00,obrien14} appear consistent with exogenous sources.

Both the Moon and Mercury provide relatively unweathered surfaces in
comparison with the Earth, so their near-surface volatile inventories
provide additional constraints on models of water delivery to the
inner solar system. Surface volatiles are not stable for
significant periods unless placed into the low temperature ``cold
traps'' provided by near-polar impact craters containing permanently
shaded regions. In order to discriminate between the various
possible sources of water \citep{moses99,crider05}, which often imply
different amounts of water being delivered, one also needs
to understand how efficiently it can migrate from the delivery
location to the cold traps \citep{but97,ong10,stewart11}. Such
models can then inform the interpretation of actual measurements of
the volatile inventories of these bodies.

Various neutron spectroscopy,
circularly polarized radar, albedo (IR, visible and UV) and impact
measurements have been 
made of the Moon and Mercury to investigate their cold traps. While
the LCROSS experiment 
\citep{cola10} found a few per cent by mass of the material in Cabeus
crater on the Moon was water ice, which is at a level consistent with neutron
spectroscopy results \citep{eke15}, there is evidence
for substantially purer water ice deposits near Mercury's poles.

A remarkable increase in the same sense circular polarized radar backscatter
cross-section was detected at wavelengths of 3.5 cm \citep{slade92,but93}, 12.6 cm
\citep{harm92,harm94,harm01} and 70 cm \citep{black10} from Mercurian
polar cold traps. One explanation for these 
radar returns was that there was multiple scattering occurring within a
low-loss medium such as water ice \citep{hapke90,hagfors97}. The high
radar backscatter regions correlated spatially with areas that have
modelled surface or near-subsurface temperatures that remain below
$\sim 100-150$\,K 
\citep{paige92,inger92,vas99}. This provided extra circumstantial
evidence to motivate the 
interpretation of the anomolous radar measurements as indicating the
presence of volatile molecules.

Instruments onboard the MESSENGER spacecraft \citep{solomon07} have
considerably increased the information available about the polar cold
traps. The deficit in 
neutron flux observed over the poles is consistent with a localised
hydrogen-rich layer extending down for tens of centimetres beneath a
10-20 cm thick layer that is less rich in hydrogen
\citep{law13}. These observations suggest that nearly pure water ice,
and not an alternative volatile such as sulphur \citep{sprague95}, 
is responsible for the 
radar features. MESSENGER's Mercury Dual Imaging System (MDIS) has
allowed an improved determination of the locations of the permanently
shaded regions, increasing the confidence with which they can be
associated with the high radar backscatter regions
\citep{chabot12,chabot13}. Albedo measurements from the Mercury
Laser Altimeter (MLA, Cavanaugh et al. \citeyear{cav07}) at 1064 nm
showed either bright or 
dark surfaces coincident with these polar cold traps
\citep{neumann13}. The distinction matched temperature model
predictions for either thermally
stable water ice at the surface (high albedo) or under a $\sim10$\,cm
thick organic lag deposit (low albedo) as noted by \citet{paige13}. 

More recently,
the sensitivity of MDIS images has allowed the imaging of permanently
shaded regions using scattered light \citep{chabot14}. The results
are similar to the MLA albedo measurements, albeit at visible
wavelengths, with anomalous reflectance regions largely coincident with
radar-bright ones. Of the surveyed areas, only Prokofiev crater has a
high albedo in MDIS images, with the other radar-bright regions
appearing anomalously dark.
The imaging in Prokofiev crater shows a
$\sim 3$\,km-wide zone that is still in permanent shade with high radar
reflectivity, but lies outside the high MDIS reflectance area. One
possible reason for this could be the presence of a stable subsurface
ice deposit that does not affect the surface
reflectance. \citet{chabot14} further note 
that the imaged, high-reflectance region displays a similar texture to
that in the sunlit region of Prokofiev, suggesting a relatively recent
ice deposition onto a previously cratered surface. \citet{deutsch16}
used a combination of MLA-derived topography and MDIS imagery to show
that radar-bright features colocate with regions of both permanent
(from MLA) and persistent (from MDIS) shadow. Furthermore, they
demonstrated that many regions of persistent or permanent shadow do not
host radar-bright deposits, and that insolation was not the
determining factor. Possible reasons for the lack of radar-bright
deposits in such apparently conducive situations were mooted to be: a
lack of radar coverage, unusually thick lag deposits hiding water ice
from the radar, and an actual lack of water ice deposits.

These various lines of evidence collectively point to the
presence of many reasonably pure water ice deposits in Mercury's polar
cold traps. Estimating the area covered by these deposits, either from
radar measurements or maps of permanent shadow if the radar misses
some \citep{deutsch16}, the 
remaining observational challenge in determining their volume 
is to measure their depth. While the neutron measurements suggest that
the hydrogen-rich layer needs to be at least half a metre
deep, for the radar results to arise due to volume scattering requires
ice at least several wavelengths thick \citep{black01,harm07}. Given the
results of \citet{black10} at 70 cm, this interpretation implies a
layer of ice at least a few metres deep. 

To place an upper limit on the thickness of possible water ice
deposits in the Mercurian polar cold traps, a few different studies
have considered the depth-diameter relations of craters.
\citet{barlow99} anticipated that subsurface ice
might lead to a softening of the terrain, as was seen on Mars
\citep{cint80,squy86}. However, they found no unequivocal evidence of
such an effect on Mercury's craters, when split as a function of
latitude. \citet{vilas05} extended this analysis, 
using the better resolution of the \citet{harm01} radar results to
focus on individual craters. With Mariner 10 imagery to determine
crater depths and diameters, \citet{vilas05} found
radar bright craters to have significantly lower depth-to-diameter
ratios than radar dark ones, to an extent that could be explained by the
presence of $\sim 900$\,m of infilling material. 
\citet{talpe12} used the MLA data
to study depth-to-diameter ratios in a sample of 537 craters poleward
of $48^\circ$N. In contrast to \citet{vilas05}, they found no evidence
for different depths, slopes or surface roughnesses for radar-bright
craters compared with their radar-dark counterparts. Ascribing the
different results to having altimetry-derived measurements,
\citet{talpe12} placed an upper limit on the depth of ice in a
$10$\,km-diameter crater of $\sim 300$\,m.

There are two orders of magnitude separating current lower and upper
limits on the depth of the ice deposits in Mercury's polar cold
traps. This paper aims to determine how the existing MLA
Gridded Data Record Digital Elevation Model (GDR DEM) can be used to
improve the constraint on the depth of these deposits in craters near
the north pole. The specific question being addressed is: are the
radar-bright regions of polar crater interiors systematically 
elevated relative to otherwise similar radar-dark parts of the
surface? 

Section~\ref{sec:data} describes the data being used in this
study. The methods for constructing the crater sample and measuring
the change in height associated with the radar-bright regions are
detailed in both section~\ref{sec:meth} and
appendix~\ref{app:meth}. Section~\ref{sec:res} contains the 
results of the analysis, and the implications are summarised in
section~\ref{sec:disc}. 

\section{Data}\label{sec:data}

This section contains descriptions of the topography and
radar backscatter cross-section data sets used to study the
Mercurian north polar craters. The topographical data were also
used to define the crater populations within which the dependence of
height with radar backscatter cross-section was studied. 

\subsection{Topographical data}\label{ssec:dem}

Data from the 11th and 15th data releases (DR11 and DR15 hereafter) of
the MLA GDR DEM, available from the Geosciences Node of
NASA's Planetary Data System
(PDS\footnote{http://pds-geosciences.wustl.edu}), have been
used. While DR11 contained a GDR at a spatial
resolution of 500\,m per pixel in a north polar stereographic
projection, DR15 included both 250\,m and 500\,m resolution DEMs. 
All of these data sets were investigated in this study, to determine the
best approach for constraining the thickness of Mercury's polar
deposits.

The MLA has absolute range uncertainties on individual altitude measurements
of better than 1m and an accuracy relative to the centre of mass of
Mercury of better than 20\,m (\citealt{sun15}; Zuber et
al. \citeyear{zuber12}).
Glitches are evident in the DR15 GDR, where the DEM height differs
systematically along particular orbital tracks by of order $\pm 100$\,m
relative to surrounding pixels. Thus, despite the improved sampling
relative to the DR11 GDR, these glitches make the DR15 GDR unsuitable for
this study and the DR11 500 m resolution option will be the
default choice for the rest of this paper.
A square region in the polar stereographic projection out to 
$\pm 1536$\,km from the north pole in both directions will be
considered. This reaches a latitude $+55^\circ$ along the coordinate
axes. The pixelated DEM has altitudes in the range $-6 < a/{\rm km} < 2.5$,
and the mean number of observations per pixel is $\sim 0.11$. This
sampling rate increases toward higher latitudes, peaking at $\sim 0.5$
observations per pixel at $+84^\circ$. The MLA GDR DEM includes
interpolated altitudes for unobserved pixels.

\subsection{Radar data}\label{ssec:radar}

The Arecibo S-band (12.6 cm) radar data from \citet{harm11}\footnote{available
at http://www.naic.edu/{\raise.17ex\hbox{$\scriptstyle\sim$}}radarusr/Mercpole/} were used to determine
the radar properties of the surface in the vicinity of Mercury's north
pole; more specifically, the $\sim0.5$\,km pixelated same-sense
circularly polarized cross-section map, $\sigma_{\rm SC}$, shown in figure
4 of \citet{harm11}. These pixels slightly oversample the instrumental
resolution of $\sim 1.5$\,km, and there may be systematic location
mismatches on 
the order of $\sim 2$\,km between this radar grid and the MLA DEM
\citep{harm11,chabot13,deutsch16}. Root mean square measurement uncertainties on
the individual 
pixel $\sigma_{\rm SC}$ values are $\sim 0.014$ and a threshold of
$0.1$ was adopted by default to separate radar-bright ($\sigma_{\rm SC}>0.1$)
pixels from radar-dark ones. This is somewhat larger than the
$4~\sigma_{\rm SC}$ level advocated by \citet{harm11}, and should
correspond to regions with thicker water ice deposits. Bilinear
interpolation is used to associate a 
same-sense radar backscatter cross-section with each MLA pixel.

\section{Methods}\label{sec:meth}

This study requires an objectively selected set of craters and a
method for assessing whether or not the radar-bright regions are
typically at a different height to the radar-dark regions within the
chosen craters. The construction of a crater catalogue from the DR11 MLA
GDR DEM, the selection of craters from this catalogue, and the method
of searching for height differences are described within the following
three subsections.

\subsection{Crater finding}\label{ssec:crat}

\begin{figure*}
\begin{center}
\includegraphics[trim=0.cm 1.cm 0cm 1.cm,clip=true,width=2.1\columnwidth]{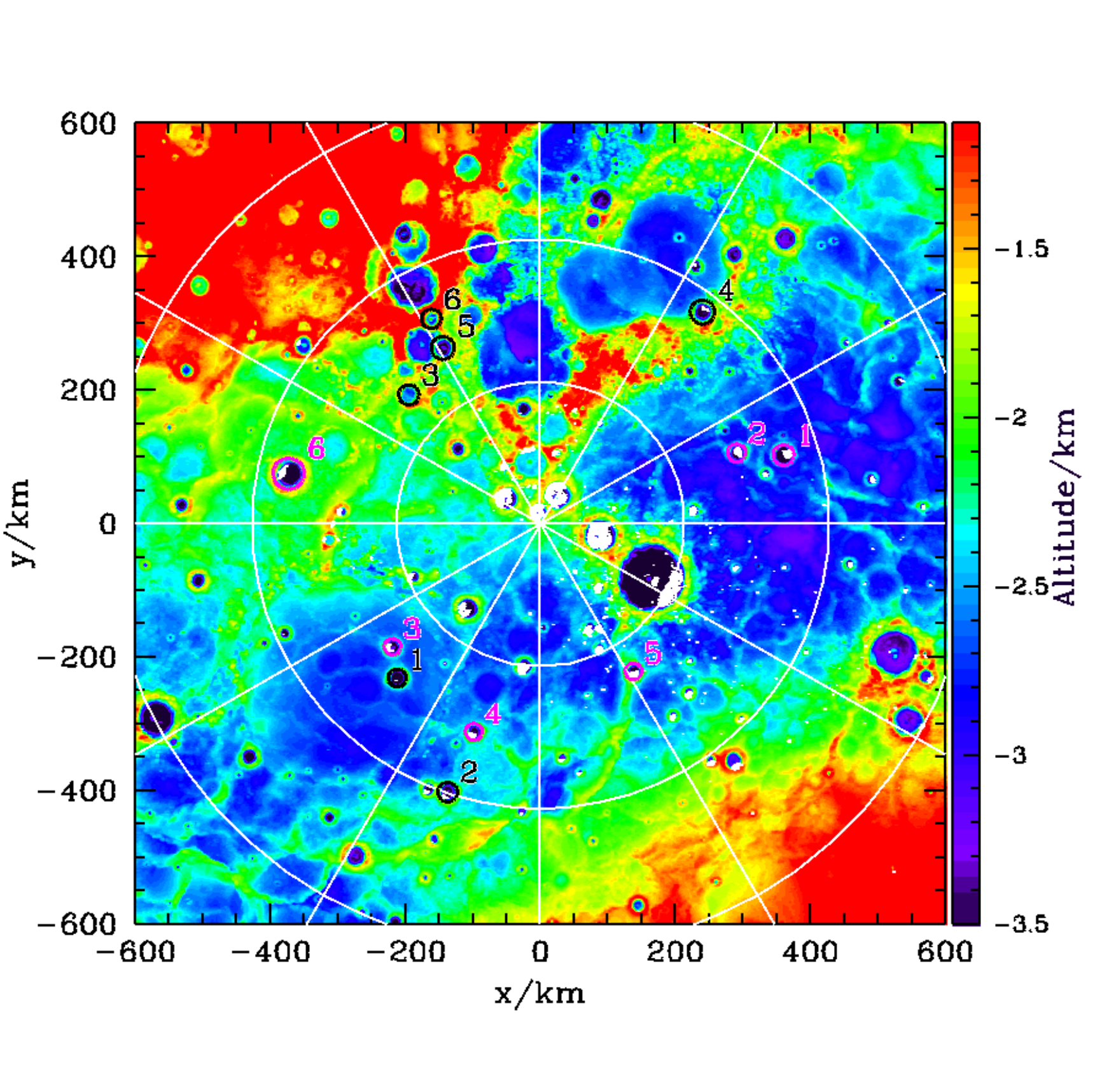}
\end{center}
\caption{DR11 MLA GDR DEM, colour-coded by height. Magenta circles
  show the radar-bright crater sample, with their identifying numbers. Black
  circles are the control sample. White points show pixels where $\sigma_{\rm
  sc}>0.1$, according to the data of \citet{harm11}. White circles
  show latitudes of $+85^\circ, +80^\circ$ and $+75^\circ$, with white
  lines every $30^\circ$ of longitude.}
\label{fig:mla}
\end{figure*}

\begin{table}
\begin{center}
\caption{Radii and locations for craters used in this
study. Longitudes and latitudes are given in degrees. Crater numbers
are shown in Fig.~\ref{fig:mla}.}
\begin{tabular}{l|c|c|c}\hline
Crater \# & Name & $r_{\rm c}$/km & latitude, longitude \\  \hline
Radar-bright & & \\
1 & Yoshikawa & 15.7 & 81.2, 106.0 \\
2 & & 13.6 & 82.7, 110.1 \\
3 & Laxness & 12.3 & 83.3, -49.9 \\
4 & Ensor & 12.6 & 82.3, -17.6 \\
5 & Carolan & 12.4 & 83.9, 31.9 \\
6 & Desprez & 23.2 & 81.1, -101.4 \\
\hline
Radar-dark & & & \\
1 & Fuller & 13.3 & 82.6, -42.6 \\
2 & Varma & 14.8 & 80.0, -18.8 \\
3 & & 14.9 & 83.6, -134.8 \\
4 & & 18.0 & 80.7, 142.9 \\
5 & & 17.1 & 82.9, -151.2 \\
6 & & 14.5 & 81.9, -152.3 \\
\hline
\end{tabular}
\end{center}
\label{tab:loc}
\end{table}

To compare the heights of radar-bright and radar-dark regions
within north polar craters requires a set of craters. As
craters containing permanently shaded areas are quite deep, one
would ideally have a set of such crater centres and their radii,
co-registered with the MLA DEM in the vicinity of the north pole. Publicly
available crater catalogues for Mercury exist, but are based on
imagery, either from Mariner 10 and the early MESSENGER flybys that do
not include much of one hemisphere near the north pole
\citep{fassett11,herrick11}, or using the full MESSENGER set of images
poleward of $+80^\circ$ \citep{deutsch16}.

Results from automated crater detection algorithms using only DEMs and
not imagery have been reported in the past for crater finding on both the Moon
\citep{luo13,xie13,li15} and Mars \citep{bue07,step09,sala10,di14},
but not yet Mercury. As the focus of this paper is on deep, symmetric
craters, a relatively simple crater detection algorithm has been
designed that should recover such features from the MLA DEM.

The details of the algorithm being used here are given in the appendix.
Briefly, the method involves filling up the DEM with virtual water
and looking for near-circular depressions associated with the
resulting puddles. When the depression stops being circular or
having enough of a circular rim as the virtual water level increases,
then a crater is defined by the final circle for which this
depression was considered to be a crater candidate.
A set of polar craters was found by applying this crater finding
algorithm to a polar subset of the DR11 MLA GDR $500$ m north pole
stereographic DEM.
This yielded 663 craters with radius, $r_{\rm c}>2$\,km, of which
266 are poleward of $+80^\circ$. For comparison, the
\citet{deutsch16} crater catalogue has 274 such craters. Most of the
larger, fresher craters are in common between the MLA-based catalogue
and that of \citet{deutsch16}. However the image-based approach
contains more craters with $r_{\rm c}<3$\,km, where the MLA
sampling makes it difficult to find circular depressions, and
does not include some of the more degraded, larger craters that the
MLA-based technique detects.

\subsection{Crater selection}\label{ssec:cratsel}

\begin{figure*}
\begin{center}
\includegraphics[trim=0.cm 6cm 0cm 4.cm,clip=true,width=2.2\columnwidth]{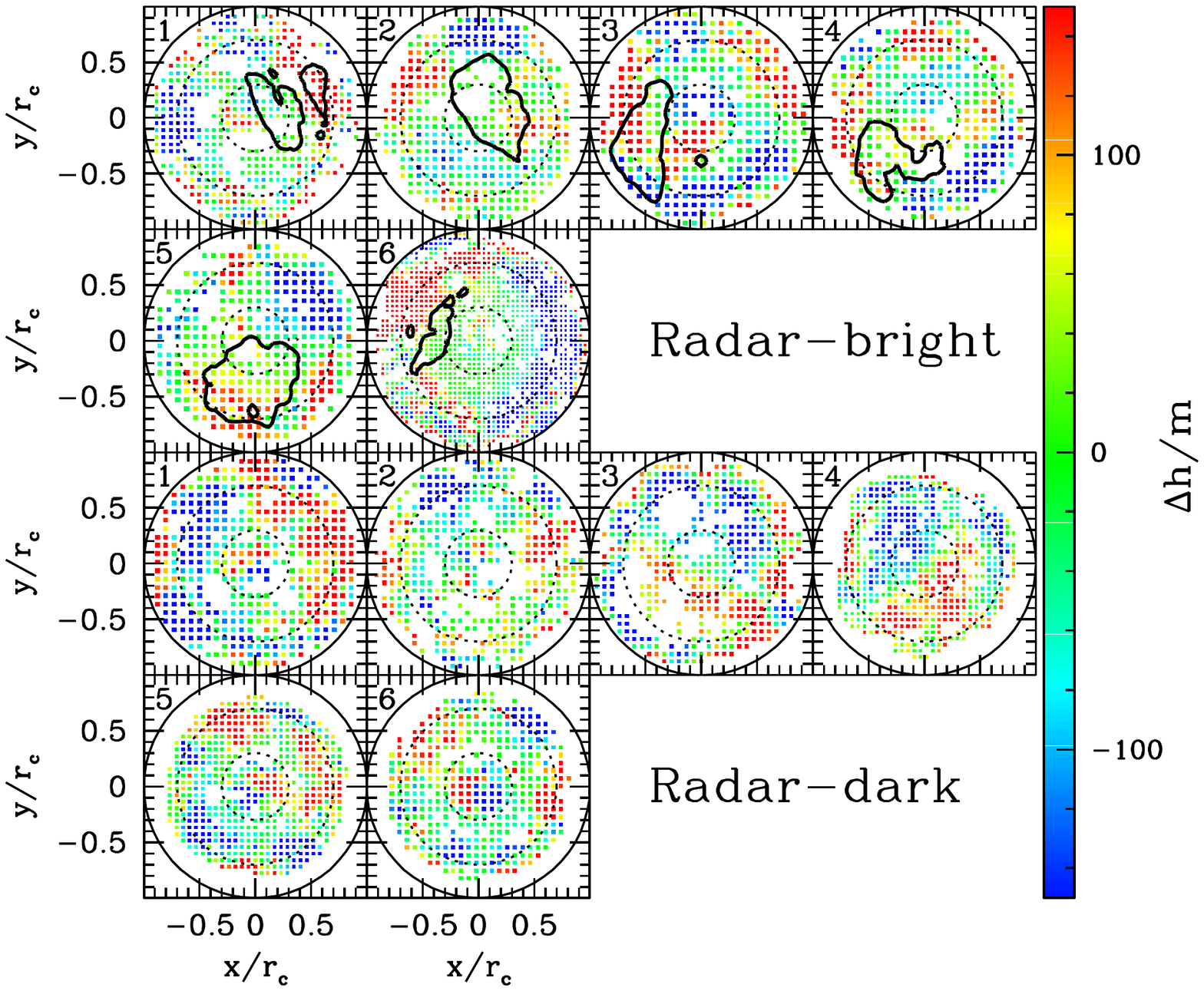}
\end{center}
\caption{Maps of height difference, $\Delta h$, for the 6 radar-bright
  craters (top two rows), with crater 1 in the top-left panel.
  The radar-bright region ($\sigma_{\rm sc}>0.1$) is delineated with
  the thick black contour. A solid black circle shows the edge of the
  crater rim at $r/r_{\rm c}=1$, and dotted circles bound the region at
  $0.3\leq r/r_{\rm c}\leq 0.7$, from which the pixels are used for
  this analysis. The lower two rows contain the 6 radar-dark
  craters that form the control sample. In all cases, only pixels
  containing at least one observation and that were placed directly into
  the puddle that contained this crater are shown.}
\label{fig:maps}
\end{figure*}

The craters to be considered need to be well-sampled by MLA
pixels. They should also be large enough that mismatches between 
radar and topography data locations are small relative to the sizes of
the craters. This puts limits on the range of latitudes and radii that
are suitable. Craters with $80^\circ<$latitude$<84^\circ$ and 
$r_{\rm c}>10$\,km, with at least $15\%$ of their 500 m pixels at
$0.3<r/r_{\rm c}<0.7$ 
containing DR11 observations are considered. Furthermore,
younger craters should be more symmetric, having had less time for
subsequent bombardment and morphological degradation. Using only
younger craters will make it easier to see the signal of any
ice-related non-axisymmetries in topography, without the additional noise
caused by extensive sub-cratering. Therefore,
craters with depth-to-diameter ratios of $d/D>0.05$ are selected.
Applying latitude, radius, MLA sampling and depth-to-diameter cuts
leads to a sample of 12 craters.

The aim is to determine, for each crater of interest, whether the
radar-bright pixels in the crater are higher or lower than the
radar-dark ones at a similar crater-centric distance, 
$r/r_{\rm c}$. In making this measurement, noise is introduced by both
steep gradients within craters not being sampled in a
similar fashion for radar-bright and radar-dark
regions and subcratering. Two additional data cuts are applied to
reduce the noise being introduced by these effects. Firstly, 
only pixels with $0.3<r/r_{\rm c}<0.7$ are included in the analysis,
removing regions with the steepest gradients caused either by central
peaks or crater walls. Secondly, only pixels that were
placed into each crater directly as the virtual water level rose are
used, to remove any subcraters from the analysis. These extra cuts
reduce the scatter in measured height differences between 
radar-bright and radar-dark regions that might otherwise act to
obscure any small changes due to the presence of an ice deposit. 

After pruning the data in this way, the remaining well-sampled craters
are placed into either the
radar-bright or radar-dark subset depending on the fraction of
radar-bright MLA-sampled pixels, $f$, in the region $0.3<r/r_{\rm c}<0.7$. 
A crater qualifies as radar-bright if $f>0.1$, whereas
radar-dark craters, which will be used as a control sample, have
$f<0.1$. These cuts 
lead to a sample of 6 radar-bright craters and 6 radar-dark ones. 
Figure~\ref{fig:mla} shows the whereabouts of the selected craters,
along with some identifying numbers, and Table~1 lists
their coordinates.
The positions of these craters in the depth-diameter plane are shown in
Figure~\ref{fig:appdD}. With the exception of Varma, which is very
near to the $+80^\circ$ boundary, all of the craters in Table~1 are
also present in the catalogue of \citet{deutsch16}.

It should be noted that no assumption is made about the presence
or otherwise of water ice in the radar-dark craters. Indeed, it is
apparent in Figure~\ref{fig:mla} that two of the ``radar-dark'' sample
have small regions where $\sigma_{\rm c}>0.1$. The presence of any
significant ice deposits in these craters would make the
control craters less axisymmetric. This, in turn, would broaden the
distribution of height differences determined from the control sample.
Hence, the confidence with which any height difference measured in
the radar-bright sample could be ascribed to the presence of a
volatile would be reduced. Thus the uncertainty derived from the
control sample will be conservatively large as a result of any bias
related to the possible presence of water ice in the control craters.

\subsection{Measuring the height difference of the radar-bright
  regions}\label{ssec:dh}

Given that craters are typically not perfectly axisymmetric
and have complicated topographies as a result of subsequent
impacts adding subcraters and degrading their rims, a statistical
approach is adopted to determine if departures from axisymmetry in the
crater topography correlate with the locations of radar-bright features.

For each selected crater, the subset of observed pixels with
$0.3<r/r_{\rm c}<0.7$ that were placed into this crater directly were
used. Pixels associated with puddles other than the one that
gave rise to the crater being considered are excluded. The typical
height as a function of cratercentric distance is determined
empirically for each crater by ordering the remaining pixels in
cratercentric distance and calculating the mean height of each
consecutive set of 10 pixels. This number is chosen because smaller
numbers of pixels produce very noisy results and larger numbers of
pixels average over a wide range in surface height due to the non-flat
shape of the crater. The results presented here are however robust to
reasonable changes in this parameter choice.

Only pixels in the small radial ranges containing both radar-bright
and radar-dark pixels are considered for further analysis. For each of
these pixels, a height difference, $\Delta h$, is defined as the
difference between its height 
and the mean height of the 10 pixels in its small radial range.
Combining pixel $\Delta h$ values from all relevant radial ranges in a
crater, the median $\Delta h$ values can be determined for the
radar-bright and radar-dark subsets of the pixels. The difference
between these values, $h_b-h_{d}$, defines the excess height of the
radar-bright region relative to the radar-dark region in that
crater. A height difference is measured for each crater, and the
distribution of these height differences can be used to address the
question:
are the radar-bright regions of polar crater interiors systematically 
elevated relative to otherwise similar radar-dark parts of the
surface? 

\section{Results}\label{sec:res}

Maps of pixel height difference, $\Delta h$, for each of the 12 
craters being considered are shown in Figure ~\ref{fig:maps}, with the
contour of high $\sigma_{\rm sc}$ 
superimposed for the 6 radar-bright cases. Within the
craters, there are regional height deviations of typically $\pm 100$ m
from axisymmetry. The deviations from axisymmetry become largest at 
$r/r_{\rm c}>0.7$, on
the steep parts of the crater walls. In some craters with central
peaks, a similar effect can be seen at $r/r_{\rm c}<0.3$. The larger
contiguous regions lacking $\Delta h$ values are depressions
other than that which gave rise to the main crater under consideration.

\begin{figure}
\begin{center}
\includegraphics[trim=1.cm 6cm 7cm 4.cm,clip=true,width=1.\columnwidth]{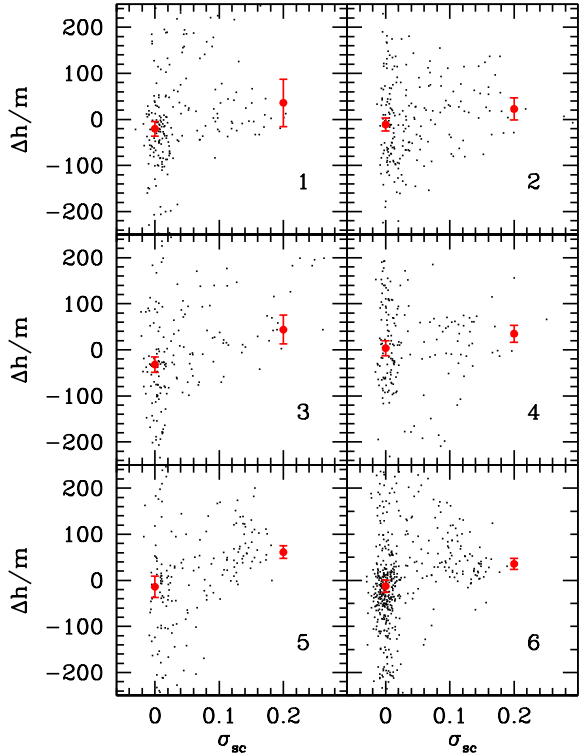}
\end{center}
\vspace{-0.2cm}
\caption{Dependence of pixel height difference, $\Delta h$, on same-sense radar
  backscatter cross-section for all pixels with $0.3\leq r/r_{\rm
  c}\leq 0.7$ within the six radar-bright craters. The crater number, as
  listed in Table~1, is shown in each panel. Red points
  show the medians of the radar-bright
  ($\sigma_{\rm sc}>0.1$) and radar-dark ($\sigma_{\rm sc}<0.1$) subsets
  of pixels. Error bars on the red points represent the $1\sigma$ errors on
  these medians, under the assumption that the individual pixel
  $\Delta h$ values have a normal distribution.}
\label{fig:cors}
\end{figure}

Figure~\ref{fig:cors} shows how pixel height difference, $\Delta h$,
depends upon $\sigma_{\rm sc}$ for each of the radar-bright
craters. The red points show the median pixel $\Delta h$ values for
the radar-bright and radar-dark pixels, and the difference between
these two values defines the statistic $h_b-h_{d}$, which measures the
excess height of the radar-bright region within each crater. For each
of the six radar-bright craters under consideration, their
radar-bright pixels are, in the median, systematically higher than the
radar-dark ones. The distribution of the height difference, $h_b-h_{d}$,
can be used to determine if this systematic difference is
statistically significant.

\begin{figure}
\begin{center}
\includegraphics[trim=0cm 4.5cm 0cm 1.5cm,clip=true,width=1.\columnwidth]{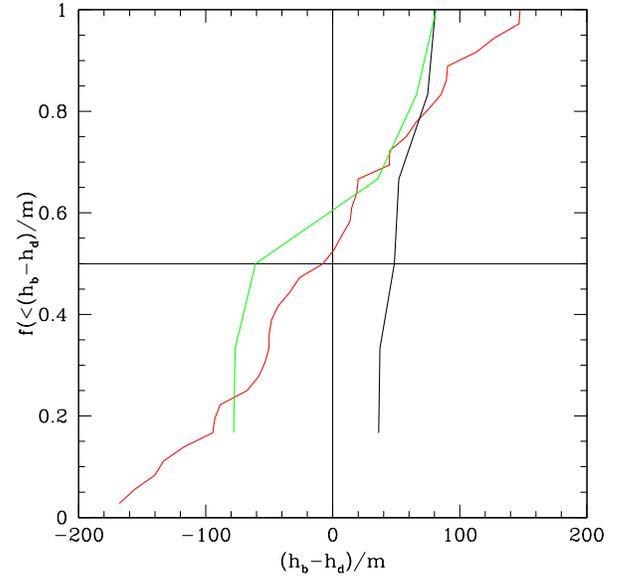}
\end{center}
\caption{Cumulative distributions of the height difference between
  radar-bright and radar-dark pixels within the radar-bright (black)
  and control sample (red) craters. All 6 radar-bright craters have
  height differences in the range $36<(h_b-h_{d})/{\rm m}<81$, hence the
  steepness of the black curve. The red curve starts at a fraction of
  1/36, this being the number of different combinations of the 6
  control craters and the 6 ``radar-bright'' region templates. A green
  curve traces the distribution of height differences for the control
  craters assuming that their ``radar-bright'' regions occupy the
  pole-facing quadrant of the craters.}
\label{fig:cdheight}
\end{figure}

For the control craters, the vast majority of pixels have 
$\sigma_{\rm SC}<0.1$. However, if the distributions of pixel 
$\Delta h$ are compared for the radar-bright and radar-dark
craters, then they are broadly similar, suggesting that the radar-dark
craters provide a suitable control sample for estimating the statistical
significance of the results.
In order to define $h_b-h_d$ for the control sample of craters, the 
$\sigma_{\rm sc}>0.1$ regions in each of the 6 radar-bright craters are
used as templates to define ``radar-bright'' regions in control
craters. This should mean that any spatial correlations in $\Delta h$ are
treated similarly between the two crater samples. Thus, the control sample
provides $6\times 6$ $(h_b-h_{d})$ values. The cumulative distributions of
the radar-bright and control crater height differences are shown in
Figure~\ref{fig:cdheight}, from which the median height difference for
the radar-bright craters is $50$\,m. 

A Kolmogorov-Smirnov (K-S) test says
that, for a null hypothesis that all craters are drawn from the same
population, there is a probability of $0.01$ that
these two distributions would be found. This is quite unlikely, in the
direction that is supportive of the possibility that there are thick ice
deposits associated with the radar-bright regions. However, the probability
is artificially lowered by the fact that the control
sample distribution has an upwards bump due to noise at
$h_b-h_{d}\approx40$\,m, where the test statistic is being determined.
The differential distribution of $h_b-h_{d}$ in the
control crater sample is well
described by a normal distribution with a standard deviation of $87$\,m, so a more
robust assessment of the median excess height associated with the
radar-bright regions in the six radar-bright craters would be
($50\pm35$)\,m. This represents a detection of a non-zero deposit
thickness at only a $\sim 1.5\sigma$ level of statistical significance.

If a threshold value of $\sigma_{\rm SC}=0.056$ had been chosen to
differentiate between radar-dark and radar-bright regions, then Varma
crater (number 2 in the radar-dark set) would switch to being a
radar-bright crater, the K-S test would return a probability of $\sim
0.09$ and the median excess height would become ($61\pm44$)\,m - again
a $\sim 1.5\sigma$ detection. This illustrates how the K-S probability
is less robust to small changes in the sample than the median height
difference. 

Also shown in Figure~\ref{fig:cdheight} is the cumulative height
difference distribution for the control sample of 6 craters, where the
``radar-bright'' region has been chosen to be the pole-facing quadrant
of the crater. No significant systematic height difference is apparent
between the pole-facing part of these craters and the remaining 3
quadrants, with the median being $\sim -10$\,m. Had these deep, polar,
radar-dark craters contained very thick ($\gsim 100$\,m) deposits on
their pole-facing slopes, then this measurement should have detected them.

The craters in the radar-bright and control samples occupy similar
regions in the crater depth-diameter diagram, as shown in
Fig~\ref{fig:appdD}. There is no significant evidence of a correlation
of height difference with any of crater depth, depth-to-diameter, number of
observations or latitude. The results are robust to reasonable
perturbations to the parameters of the crater-finding algorithm,
because it tends always to find the same craters and the pixels that
are excluded from the analysis are defined by the topography
itself. Changing the selection of craters used to perform the test
makes a larger difference to the results. By including smaller, less
well-sampled craters, the significance of the difference between the
radar-bright and control samples decreases. Similarly, extending the
sample to lower or higher latitudes, where the craters are less
well-sampled by MLA measurements also makes the distribution of height
differences of the radar-bright sample look more like that of the
control sample.

\section{Discussion}\label{sec:disc}

From this study, the typical excess thickness associated with the
radar-bright regions in 6 north polar craters with diameters exceeding
$20$\,km is ($50\pm35$)\,m. While this does not represent a
statistically significant measurement of a non-zero height increase in
the radar-bright regions in polar cold traps, it does provide an
upper limit of $\sim 150$\,m on the depth of the typical
ice deposit that may be associated with these regions. This is a
factor of $\sim 2$ lower than that which was previously available
\citep{talpe12}. 

Given the typical, not ice-related, undulations in the surface within
craters, it is not feasible to relate a localised change in
height within any particular crater with the presence of a thick
deposit of water ice. These departures from axisymmetry, even within
fresh craters, can be up to $\sim 100$\,m in amplitude. Consequently,
improvements in range measurement accuracy, which is already much smaller
than this, will not have a significant impact. More important would be
both a more accurate positional matching between radar and DEM data sets,
which would lead to reduced scatter in pixel $\Delta h$ values caused
by positional mismatches, 
and a denser sampling of the topography. These would permit a reliable
extension of the technique developed here to include smaller craters
containing radar-bright features. With a larger sample of craters, the
difference between radar-bright and control craters could be more
accurately determined. To some level, improved laser altimeter
sampling is provided by the MLA DR15 data although, as discussed
earlier, there are presently some non-negligible glitches in the DR15
GDR that stymie this approach.

The BepiColombo Laser Altimeter (Thomas et al. \citeyear{thomas07}) is
scheduled to map 
Mercury's surface in the next decade and will have a polar orbit with
along-track resolution of $\sim 250$\,m. At the end of the mission,
the cross-track resolution should be better than $\sim 1$\,km more than
$80^\circ$ from the equator. If the elliptical orbit sees BepiColombo
flying low over the south pole, then this might double the number of
available craters; otherwise, the anticipated lateral sampling will
not differ greatly from that provided by the MLA.

Taken at face value, if a depth of $50$\,m of water ice were typical
of all the radar-bright regions near the
north pole of Mercury, then this would correspond to a total mass of
water ice deposited near Mercury's poles of $\sim 
10^{15}$\,kg, assuming that the south pole contains a similar
quantity, the water ice density is $10^3$\,kg\,m$^{-3}$ and using a
value of $\sim 10,000$\,km$^2$ for the total north pole radar-bright
area \citep{harm11}. If radar-dark permanently shaded regions also
host deposits of water ice, then this value would represent an
underestimate of the total mass present. Even so, this still
amounts to more water ice than could feasibly be delivered to Mercury by
micrometeorite bombardment, Halley-type comets or asteroids, according
to the estimations of \citet{moses99}. However, the value lies within
their predicted range for water delivery from Jupiter-family
comets. Given the uncertainty in the actual typical measured depth, it
would be premature to rule out any of the alternative delivery
sources on the basis of the results presented here. 
Furthermore, there are considerable uncertainties in the
micrometeorite flux reaching Mercury, with recent studies by
\citet{borin09,nesvorny10} and \citet{brucksyal15}
finding values that are respectively $\sim 60, 30$ and $10$ times
those assumed by \citet{moses99}.
However, this
analysis does exclude the possibility of the 
total water ice deposits exceeding $\sim 3\times 10^{15}$\,kg,
provided that the craters studied have ice depths typical of other
regions hosting deposits.

\section*{Acknowledgments}
VRE acknowledges helpful discussions with Adrian Jenkins and Wenzhe Fa.
VRE was supported by the STFC rolling grant ST/L00075X/1. 
This work used the DiRAC Data Centric system at Durham University,
operated by the Institute for Computational Cosmology on behalf of the
STFC DiRAC HPC Facility (www.dirac.ac.uk). This equipment was funded
by BIS National E-infrastructure capital grant ST/K00042X/1, STFC
capital grants ST/H008519/1 and ST/K00087X/1, STFC DiRAC Operations
grant ST/K003267/1 and Durham University. DiRAC is part of the
National E-Infrastructure.

\def\jgr{J. Geophys. Res. }
\def\grl{Geophys. Res. Lett. }
\def\nat{Nature }
\def\mnras{Mon. Not. R. Astron. Soc. }
\def\icarus{Icarus}
\def\aap{A\&A}
\def\pasp{Publ. Astron. Soc. Pac. }

%

\bibliographystyle{mn2e}

\bibliography{mybib}

\begin{appendix}

\section{Crater-finding algorithm}\label{app:meth}

In designing an automated crater-finding algorithm, the main challenges
arise because craters become degraded and often overlap
with one another. The algorithm needs to allow for the possibilities that
some craters can lie entirely within other craters, and that some may only
retain a small fraction of a circular rim. The approach taken here is
motivated by the expectations that craters will include a local
minimum in the topography, which is not inevitably the case but should
be true in the vast majority of cases, and have
at least some part of a near-circular edge.

The crater-finding algorithm comprises two main stages, the first of
which involves gradually filling up the DEM with virtual water and
creating a tree-like structure of interconnected puddles. The second
stage searches through the sets of resulting puddles looking for
near-circular depressions. It will be assumed that any
near-circular depression is a crater, although this need not be the
case. These two operations are described in this appendix.

\subsection{Creating the puddle tree}\label{apps:pud}

A set of virtual water levels spaced by $\delta a=1$\,m were used
gradually to fill up the terrain. At first a single, pixel-sized puddle was
present. Subsequently other puddles appeared and merged together until a
single large puddle covered the entire domain.

For each water level, a
two-dimensional friends-of-friends algorithm \citep{davis85} was run to find
distinct puddles. This involves linking together wet pixels,
i.e. those at altitudes beneath the virtual water level, with their
wet neighbours, including diagonal links. If two or more puddles at
one level are linked together at the next, when the water level increases by
$\delta a$ and creates more wet pixels, then the smaller progenitor
puddles are noted as having merged into the larger/largest progenitor,
which retains its identity. In this way a tree-like structure of
puddles is built up. This tree can be traversed in order to track the
varying shape of any puddle as the water level is increased.

Applying this friends-of-friends algorithm to the 500\,m MLA GDR DEM north
polar stereographic projection from DR11 leads to the identification
of 362248 distinct puddles. These puddles are present for at least one
of the 8453 virtual water levels. Had the number of levels been halved
by choosing $\delta a=2$\,m, then the number of distinct puddles would
have decreased only slightly to 344245.

\subsection{Finding near-circular depressions using the puddle tree}

Each puddle that appears as the virtual water level increases can give
rise to many craters. It is also possible that it will not host any craters.
The second part of the crater-finding algorithm involves finding
near-circular puddles, tracking how these crater candidates evolve as
the virtual water level 
rises, and logging them as craters at the level before they cease to
have a sufficiently detectable near-circular edge. This subsection
defines what it means to be near-circular and how the crater
candidates are treated if their host puddles merge into other puddles.

At a given virtual water level, a puddle is defined as a set of
pixels. Puddles are split into ``main'' puddles, which have yet to 
merge into a larger pre-existing puddle as the virtual water level
rises, and ``progenitor'' puddles, which have already merged into a
larger puddle. The 500 m sampling of the DEM limits the ability
to determine how circular small puddles are, so only puddles
containing an area of wet pixels $A\geq\pi r_{\rm c,min}^2$ are
considered as potential hosts of crater candidates. For the results
presented here, the choice $r_{\rm c,min}=2$\,km has been used,
for which the diameter of the crater will be sampled with 8
pixels. Smaller puddles are too poorly sampled to determine if they
are sufficiently circular.
At every virtual water level, every main puddle present is
assessed to determine if it contains a crater candidate. All crater
candidates from progenitor puddles are also tracked to determine if
they remain crater candidates. The centres and radii of crater
candidates are updated.

There are three distinct methods used to determine if a crater
candidate exists or remains in each sufficiently big main puddle, the
second and third of which are used for crater candidates in progenitor
puddles.

\noindent
1) For main puddles only, all pixels are used to determine 
$I=\pi \langle r^2\rangle/A$, where $\langle r^2 \rangle$ represents
the mean-squared 
separation of the constituent pixels from their mean location. If
$I < I_{\rm max}$ then this puddle is sufficiently circular to be a
viable crater candidate. Small fluctuations of $I$ near to 
$I_{\rm max}$ can lead to multiple detections of what is essentially
the same crater. To suppress the frequency of such events, any crater
candidate at the previous, lower level, remains viable provided that
$I < I_{\rm max}+\delta I$. For this work, $I_{\rm max}=0.53$ and 
$\delta I=0.02$. The candidate crater centre and radius, defined as
$r_{\rm c}=\sqrt{A/\pi}$, are updated at each level. Also, if a main puddle
will become a progenitor puddle at the next level and is not a crater
candidate, then its centre and ``radius'' are stored.

If method (1) has failed to find a crater candidate for a sufficiently
big main puddle,
or a crater candidate in a progenitor puddle is being tracked from the
previous level, then
the following procedure is followed to hunt for a crater candidate.

\noindent
2) The set of dry perimeter pixels around the main puddle are found, not
   including diagonal steps. Near-circular arc is defined as the total area,
   $A_{\rm per}$ of perimeter pixels whose centres lie in the range
   $[r_{\rm c}-\delta r/2,r_{\rm c}+\delta r/2]$ 
   from the crater candidate centre found at the previous level, where
   $\delta r=1.0$\,km is used here. If 
   $A_{\rm per}>f_{\rm rim}2\pi r_{\rm c} \delta r$, where $f_{\rm rim}=0.4$, then
   a potential crater candidate exists. For
   just-merged progenitor puddles that contained no crater candidate at
   the final level when they were main puddles, the near-circular arc
   pixels are allowed to lie in the range $[0,{\rm max}(2r_{\rm
   c},r_{\rm c}+5\delta r/2)]$.

If this method finds sufficient near-circular arc, then the potential
crater candidate centre and radius are determined using the perimeter pixels that
comprise the near-circular arc. This is done by taking up to $10^6$
distinct tuples of these pixels and determining the centres and radii
of the circles they define. The centres that lie within a distance 
$2f_{\rm shift}r_{\rm c}$ of the previous centre, where $f_{\rm shift}=0.1$,
are retained and a shrinking circles algorithm is applied to determine
the potential crater candidate centre. This involves iteratively calculating
the mean position of 
the remaining centres then reducing the radius within which centres
are included by $5\%$ before repeating. When no more than $30$ centres
remain, their mean position is
returned as the centre of the potential crater candidate. The
radius is taken to be the median of the radii associated with the
remaining centres. If the centre of the potential crater candidate
lies within a distance $f_{\rm shift}r_{\rm c}$ of the previously
calculated centre, and the radius is no more than $10\%$ larger than
the previously calculated radius, then this potential crater candidate
is deemed to be a crater candidate. Otherwise, assuming that a crater
candidate did exist at the preceding virtual water level, one final
method is attempted to try to locate a viable crater candidate.

\noindent
3) It is possible that the perimeter of dry pixels around a puddle in
   the vicinity of a crater candidate does not enclose its centre. For
   example, if the crater rim is irregular in height and only a subset
   locally pokes above the virtual water level. In this case, rather
   than using the puddle perimeter as was done in method (2),
   potential dry rim pixels are sought a distance $[r_{\rm c}-\delta
   r/2,r_{\rm c}+\delta r/2]$ away from the previous crater candidate
   centre. If the area of dry rim satisfies 
   $A_{\rm rim}>f_{\rm rim}2\pi r_{\rm c} \delta r$, then a viable
   crater candidate remains and its radius is updated to be the mean
   radius of the dry rim pixels from the previous centre. The centre
   is not updated in this method.

If a crater candidate existed at the previous level and it ceases to exist, then its
last acceptable centre and radius are logged as a crater.

\subsection{Results}\label{apps:res}

\begin{figure}
\begin{center}
\includegraphics[trim=0cm 5cm 0cm 1cm,clip=true,width=0.95\columnwidth]{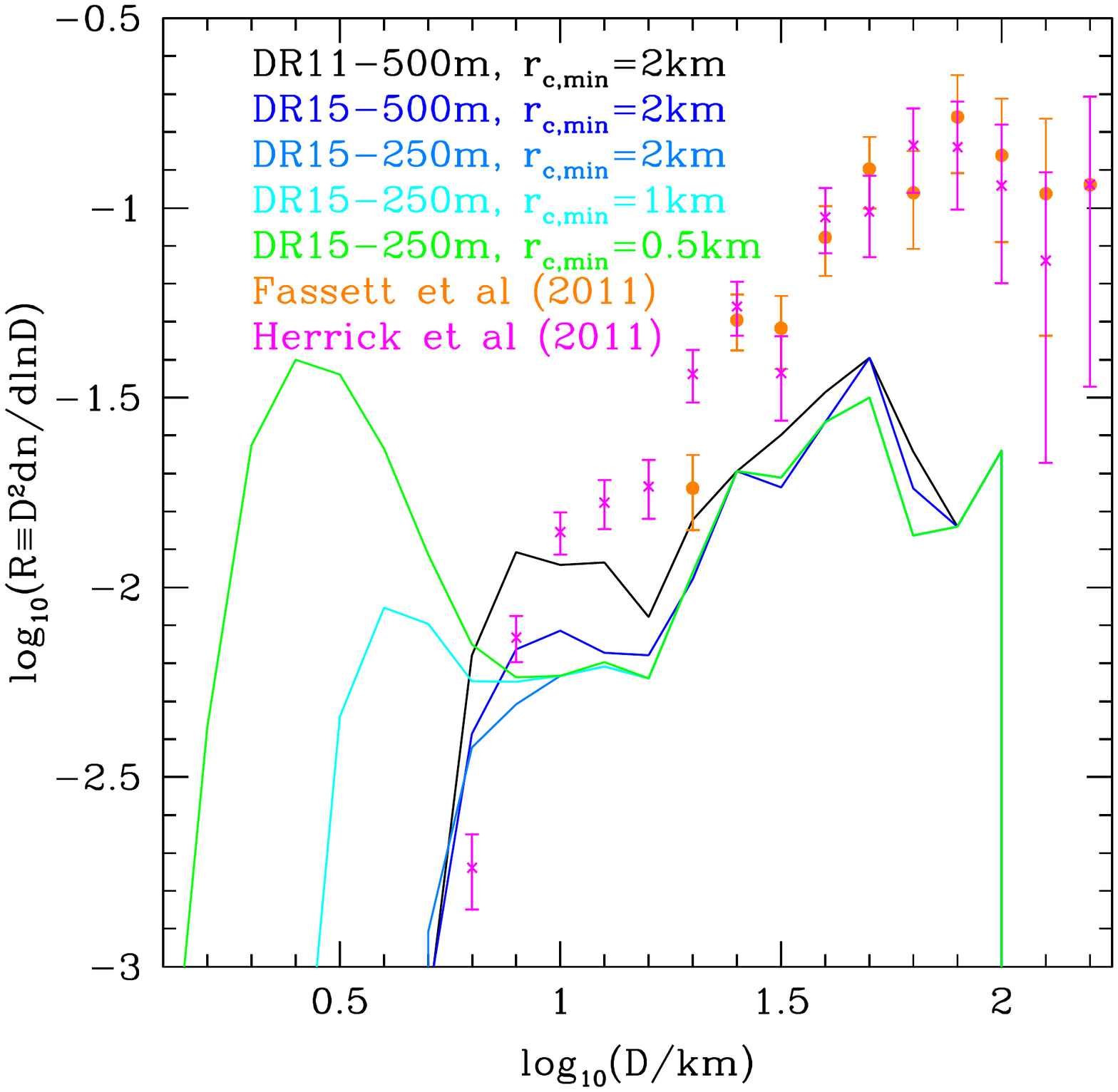}
\end{center}
\caption{The abundance of craters as a function of crater diameter,
  $D$ for the region $|x|,|y|<800$\,km from the pole in the polar
  stereographic projection. The standard $R$-plot is shown
  \citep{crat79}, where $n$ 
  represents the number of craters per unit area. Curves represent the
  crater abundances from applying the algorithm described in this
  appendix to the different GDR data releases, at different pixel
  resolutions, with different minimum crater candidate sizes, as shown
  in the legend. The points with error bars are using craters from the
  catalogues of \citet{fassett11} and \citet{herrick11} in the same region.}
\label{fig:appnofr}
\end{figure}

\begin{figure}
\begin{center}
\includegraphics[trim=0cm 5cm 0cm 1cm,clip=true,width=0.95\columnwidth]{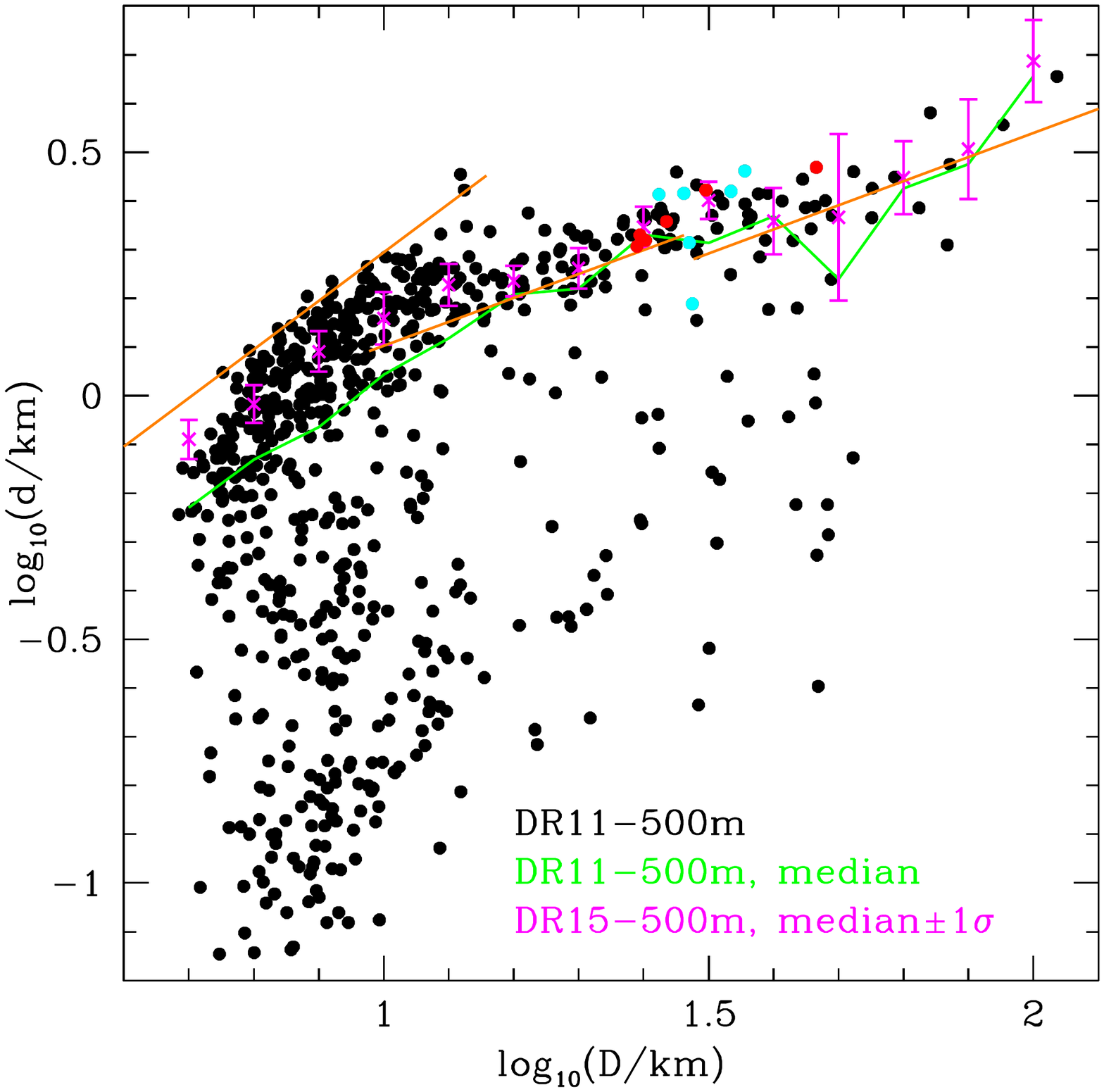}
\end{center}
\caption{The depth-diameter relation for the recovered craters. Black
  filled circles show all 663 DR11 craters found using $r_{\rm c,min}=2$\,km,
  and the green line traces the median depth as a function of
  diameter. The magenta crosses are the corresponding medians for the
  663 DR11 craters, with error bars showing the errors on the median
  depths under the assumption of Gaussian distributed depths at each
  diameter. In order of increasing diameter, the three orange lines
  represent the relations for simple craters, immature and mature
  complex craters determined by \citet{pike88}. Red and cyan filled
  circles show the 6 radar-bright and 6 radar-dark craters respectively
  used in this study.}
\label{fig:appdD}
\end{figure}

Applying the algorithm described above to the MLA GDR DR11
leads to a database of 663 craters with radii $\geq 2$\,km. The
crater counts as a function of crater diameter are shown in
Fig.~\ref{fig:appnofr}, in comparison with the results from
\cite{fassett11} and \citet{herrick11}. It is important to note that
the catalogues provided by \cite{fassett11} and \citet{herrick11} are
derived from MESSENGER and Mariner 10 images 
that only adequately cover approximately half of the area being
used. As a consequence, the abundances shown in Fig.~\ref{fig:appnofr}
have been multiplied by $2$ to account for this survey mask. This
factor of two is approximately the incompleteness in the catalogue
derived here for crater sizes in the range 
$10\lsim r_{\rm c}/{\rm km}\lsim 25$. The incompleteness is greater for
larger craters. This reflects the difficulty of automatically finding
very degraded craters. However, the new set of craters extends down to
smaller radii than were previously available, and covers the entire
range of longitudes near to the north pole.

The various different
lines in Fig.~\ref{fig:appnofr} show the effect of using data from
different releases (DR11 and DR15), with different pixel sizes (250\,m
and 500\,m) and using different values for the parameter $r_{\rm c,min}$.
The difference between green and cyan curves is that crater candidates
are allowed to be smaller in the former case. Given that these 
$r_{\rm c,min}$ values are close to the pixel size in the DEM, it is
difficult to determine accurately whether or not a puddle is nearly
circular enough to host a crater candidate. From the convergence of
the curves, it appears that the results are robust for craters with
$r_{\rm c}\geq 3$\,km, which is approximately 10 times the pixel
size. The blue curve shows how using the 500\,m-resolution DEM, rather
than the 250\,m DEM alters the crater abundances. A more significant
change is seen when using the DR11 GDR. Almost twice as many 
$r_{\rm c}\sim 10$\,km craters are found. The results converge for larger
craters. While the DR11 GDR is less well-sampled than DR15, the newer
release contains glitches, where particular orbits are $\sim\pm
100$\,m different in height than the surrounding measurements, so it
is not immediately clear which of these data sets is to be preferred.

Another interesting way to characterise the craters is through their
depth and diameter. These are shown in Fig.~\ref{fig:appdD}, where the
crater depth is simply defined as the height
difference between the minimum and maximum altitude pixels with
centres lying within $r_{\rm c}$ of the crater centre. The set of DR11
craters includes significantly more shallow, small-diameter craters
than for the DR15 case, pulling the median depth down for this
case. In comparison with the results of \citet{pike88}, the
automatically found craters are typically slightly shallower than his
simple crater sample and deeper than his complex craters. Given the
different methods for finding craters and measuring diameters and
depths, the results are similar. 

In summary, the automated DEM-based crater-finding algorithm presented
here is finding approximately half the larger craters present in the
\citet{fassett11} and \citet{herrick11} samples, as well as plenty of
smaller ones. The 
missing craters are predominantly those that are less well-defined due
to degradation by subsequent bombardment.
For the purpose of the study presented here, the fact that the
algorithm is finding the deep, fresh craters is the most important
point. 

\end{appendix}


\end{document}